\documentclass[conference]{IEEEtran}
\IEEEoverridecommandlockouts  

\usepackage{cite}
\usepackage{amsmath,amssymb,amsfonts}
\usepackage{graphicx}
\usepackage{textcomp}
\usepackage{xcolor}
\usepackage{booktabs}
\usepackage{array}
\usepackage{tabularx}
\usepackage{url}
\usepackage{flushend}

\newcolumntype{Y}{>{\raggedright\arraybackslash}X}
\newcolumntype{P}[1]{>{\raggedright\arraybackslash}p{#1}}

\makeatletter
\newcommand{\linebreakand}{%
  \end{@IEEEauthorhalign}
  \hfill\mbox{}\par
  \mbox{}\hfill\begin{@IEEEauthorhalign}
}
\makeatother

\begin{document}



\title{Quality Assessment of 3D Gaussian Splatting: Distortions, Benchmarks, and Open Challenges\\
\thanks{This work was supported in part by the National Natural Science Foundation of China (Grant No.62571160), the Engineering Technology R\&D Center of Guangdong Provincial Universities (Grant No.2024GCZX004).}
}
\author{
\IEEEauthorblockN{1\textsuperscript{st} Shuai Liu}
\IEEEauthorblockA{
\textit{College of Applied Technology}\\
\textit{Shenzhen University}\\
Shenzhen, China\\
2205414001@email.szu.edu.cn
}
\and
\IEEEauthorblockN{2\textsuperscript{nd} Binqiang Liu}
\IEEEauthorblockA{
\textit{College of Electronics and}\\
\textit{Information Engineering}\\
\textit{Shenzhen University}\\
Shenzhen, China\\
2410044032@mails.szu.edu.cn
}
\and
\IEEEauthorblockN{3\textsuperscript{rd} Qingyu Mao}
\IEEEauthorblockA{
\textit{College of Electronics and}\\
\textit{Information Engineering}\\
\textit{Shenzhen University}\\
Shenzhen, China\\
qingyu.mao@outlook.com
}
\linebreakand
\IEEEauthorblockN{4\textsuperscript{th} Jiacong Chen}
\IEEEauthorblockA{
\textit{College of Applied Technology}\\
\textit{Shenzhen University}\\
Shenzhen, China\\
2210434045@email.szu.edu.cn
}
\and
\IEEEauthorblockN{5\textsuperscript{th} Yongsheng Liang\textsuperscript{*}}
\IEEEauthorblockA{
\textit{College of Applied Technology}\\
\textit{Shenzhen University}\\
Shenzhen, China\\
liangys@szu.edu.cn
}
\and
\IEEEauthorblockN{6\textsuperscript{th} Youneng Bao\textsuperscript{*}}
\IEEEauthorblockA{
\textit{College of Electronics and}\\
\textit{Information Engineering}\\
\textit{Shenzhen University}\\
Shenzhen, China\\
baoyn@szu.edu.cn
}}

\maketitle

\begingroup
\renewcommand\thefootnote{}\footnotetext{*Corresponding authors: Yongsheng Liang and Youneng Bao}
\endgroup

\begin{abstract}
3D Gaussian Splatting (3DGS) has become a practical scene representation for real-time novel-view rendering, compression, and immersive content delivery. However, its quality assessment still largely follows rendered-view proxy protocols that sample camera poses, render images or videos, and apply inherited image and video quality metrics. While convenient, this practice does not fully capture 3DGS-native degradations that originate from Gaussian primitive distributions, splatting and visibility behavior, and trajectory-dependent artifacts. This survey reviews recent 3DGS quality assessment studies from four perspectives, covering distortion characteristics, subjective benchmarks, objective metric reliability, and emerging directions for native 3DGS evaluation. Across the literature, we find that different benchmarks construct different notions of quality, and that metric effectiveness is highly protocol dependent, varying with distortion sources, stimulus formats, and view and trajectory sampling. Finally, we summarize open challenges and outline practical guidelines for building more comparable and representative 3DGS-QA evaluations.
\end{abstract}


\section{Introduction}

Neural scene representation is increasingly evolving from purely implicit radiance fields to explicit, renderable scene primitives that better support efficient rendering, compression, transmission, and editing. As a representative milestone in this direction, 3D Gaussian Splatting (3DGS) represents scenes with anisotropic Gaussian primitives parameterized by opacity and view-dependent appearance, inheriting ideas from splatting-based rendering while departing from the implicit neural field formulation popularized by NeRF~\cite{zwicker2001,mildenhall2020,kerbl2023}. As a result, 3DGS has become a practical foundation for compression and streaming of dynamic content, extended reality, digital twins, and other interactive media pipelines~\cite{fei2024survey}.

Evaluation methodology for 3DGS has not matured at the same pace. Most current protocols sample a finite set of camera poses, render images or videos, and then report PSNR, SSIM, or learned image/video quality assessment (IQA/VQA) scores. Such rendered-view assessment is indispensable because it is easy to compare, and aligns well with many rate–distortion settings. However, it evaluates a 3DGS model only through a particular view-sampling policy, so high scores on the selected views do not necessarily imply stable primitive geometry, cross-view consistency, temporal smoothness, or a satisfactory free-viewpoint experience.

This mismatch is structural rather than merely numerical. Artifacts in 3DGS are often sparse, localized, and view-dependent. Distortions may occupy only a small region of a frame, yet they can dominate human perception when a user navigates the scene. Conversely, a rendering can achieve strong full-reference image scores while masking unstable depth, abnormal primitive distributions, or failures that surface only from untested viewpoints. In this sense, what we evaluate is not just a set of rendered images, but a scene representation experienced through views and trajectories.
\begin{figure*}[!t]
\centering
\includegraphics[width=0.92\textwidth]{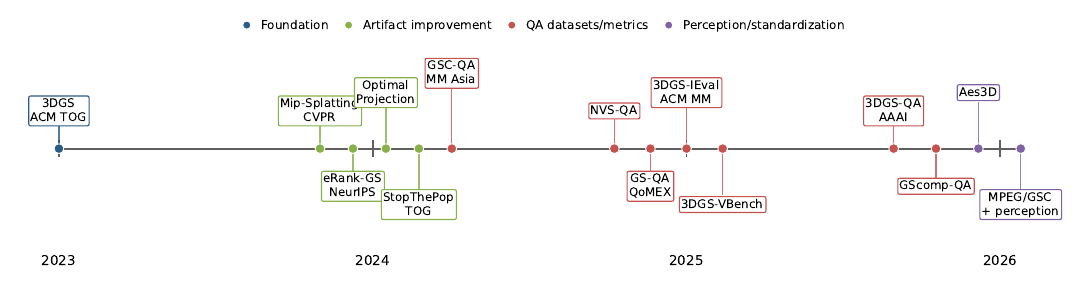}
\caption{Representative milestones in 3DGS-QA. The timeline highlights how the field has moved from rendered-view fidelity reporting toward artifact-specific, subjective, perceptual, compression-oriented, aesthetic, and standardization-aware evaluation.}
\label{fig:timeline}
\end{figure*}

Recent benchmarks make this tension measurable. Controlled 3DGS quality-assessment studies report weak correlations between conventional full-reference metrics and human judgments on deliberately designed 3DGS distortions, whereas compression-oriented image/video benchmarks often favor learned IQA/VQA models under fixed stimulus protocols~\cite{wan2026qa,xing2025ieval,xing2025vbench,martin2026gscompqa}. Multi-distance and uncertainty-aware datasets further show that subjective quality varies with input resolution, the number of input views, camera distance, and initialization quality~\cite{chen2025mugsqa}. Large-scale perceptual preference studies further quantify this mismatch. In a blind pairwise preference study with 428 participants and 39,320 comparisons, a Wasserstein-distortion-regularized 3DGS variant was preferred 2.3$\times$ more often than a model trained with the conventional L1 and SSIM objective, despite lower PSNR/SSIM in some comparisons~\cite{ozyilkan2026dropin}. Adjacent novel-view synthesis and NeRF quality-assessment studies reach a similar conclusion from a different representation. Perceptual quality depends on view sampling, trajectory design, and temporal behavior, not only on single-image fidelity~\cite{martin2023nerfqa,liang2024nvsqa,qu2024nerfnqa}. These findings are not contradictory. Rather, they reflect that each benchmark operationalizes a different notion of quality. Fig.~\ref{fig:timeline} summarizes this evolution in 3DGS quality assessment (3DGS-QA), from rendered-view fidelity reporting toward artifact-specific, subjective, perceptual, and standardization-aware evaluation.

Motivated by these observations, we synthesize recent work into a compact critical survey of 3DGS-QA. Instead of introducing yet another dataset or metric, we make explicit the evaluation assumptions that are often hidden behind a single rendered-view score. Our discussion is organized around four questions: which distortions are specific to 3DGS, what subjective benchmarks actually measure, when objective metrics are reliable, and what a native 3DGS quality protocol should include. The contributions of this work are as follows:

\noindent$\bullet$ We organize 3DGS-specific distortions into a representation-aware taxonomy and relate each distortion category to its implications for quality assessment.

\noindent$\bullet$ We compare 3DGS-QA datasets along methodological axes rather than scale alone, highlighting stimulus design, rating protocols, scene and trajectory coverage, reliability analysis, and the specific quality construct each benchmark targets.

\noindent$\bullet$ We reinterpret objective metrics as conditional evidence and outline a native evaluation framework that combines rendered observations, primitive-level diagnostics, view-sampling controls, temporal probes, and human validation.

\section{3DGS-Specific Distortions}

A credible 3DGS-QA protocol must consider the representation. A Gaussian primitive is neither a surface element in a mesh nor a point sample in a point cloud. It encodes position, anisotropic scale, rotation, opacity, and appearance coefficients, and its perceptual impact depends on projection, blending, depth ordering, filtering, and the viewing trajectory. Consequently, the same visible artifact may stem from geometry, rendering, optimization, compression, or view sampling.

\begin{table*}[!t]
\caption{Representative 3DGS-specific distortions and their implications for quality assessment.}
\label{tab:distortions}
\centering
\scriptsize
\setlength{\tabcolsep}{3.4pt}
\renewcommand{\arraystretch}{1.08}
\begin{tabularx}{\textwidth}{P{0.13\textwidth} P{0.25\textwidth} P{0.22\textwidth} P{0.15\textwidth} Y}
\toprule
Distortion & Mechanism & Visual symptom & Representative work & Quality assessment implication \\
\midrule
Needle-like & Extreme anisotropy or collapse in Gaussian scaling & Thin streaks or sharp elongated structures & eRank-GS~\cite{erankgs2024} & Small image area can still be perceptually salient. \\
Floating & Erroneous primitives detach from the true surface & Floaters, fragments, or ghost structures & EFA-GS~\cite{efags2025} & Visibility varies by viewpoint and motion. \\
Popping & Depth sorting, visibility, or splatting order changes across views & Abrupt flicker or discontinuity & StopThePop~\cite{radler2024stopthepop} & Isolated frames can miss the event. \\
Cloud-like blur & Oversized projected splats or projection approximation error & Fuzzy regions and cloudy surfaces & Optimal Projection~\cite{huang2024optimal} & Pixel fidelity may hide structural causes. \\
Aliasing & Insufficient multi-scale filtering of high-frequency content & Distance-dependent jaggies or shimmer & Mip-Splatting~\cite{yu2024mipsplatting} & Quality depends on scale and camera pose. \\
Controlled distortions & Sparse views, under-training, point downsampling, noise, or color perturbation & Systematic reconstruction degradation & 3DGS-QA~\cite{wan2026qa} & Enables controlled subjective and metric analysis. \\
\bottomrule
\end{tabularx}
\end{table*}

The literature on 3DGS-Specific distortions already reflects this diversity, although it is typically framed as rendering or optimization research rather than quality assessment. Anti-aliasing, view-consistent rendering, projection analysis, rank regularization, and frequency-aware artifact suppression each address a different 3DGS failure mode~\cite{yu2024mipsplatting,radler2024stopthepop,huang2024optimal,erankgs2024,efags2025}. These works identify failure modes that a 3DGS-QA protocol should be able to reveal. The taxonomy in Table~\ref{tab:distortions} therefore serves as a bridge between rendering research and perceptual evaluation.

A defining property of these distortions is their nonuniform perceptual footprint. A needle-like~\cite{erankgs2024} or floating~\cite{efags2025} streak may occupy only a few pixels and contribute little to average error, yet its semantic implausibility and motion-dependent visibility can be highly detrimental to perceived quality. Cloud-like blur~\cite{huang2024optimal} may spread over a large region but has a different perceptual signature, as it primarily degrades surface sharpness rather than introducing discrete spurious structures. Aliasing~\cite{yu2024mipsplatting} and popping~\cite{radler2024stopthepop} are likewise trajectory-dependent, varying with camera motion, scale, and depth ordering. As a result, a single rendered-view score can conflate qualitatively different perceptual cues.

This pattern is rooted in the 3DGS representation. Small primitive-level errors can produce locally abrupt artifacts that barely change PSNR/SSIM yet strongly affect perception. This asymmetry also helps explain why metric behavior varies across datasets. Therefore, 3DGS calls for a dedicated perceptual quality assessment framework, rather than a direct transplantation of conventional IQA/VQA protocols.

\section{Subjective Benchmarks and Dataset Coverage}

Subjective benchmarks provide the human evidence used to calibrate objective quality assessment. In 3DGS, a benchmark specifies more than a collection of distorted models. It also fixes the scene class, reconstruction source, view and trajectory sampling, stimulus format, rating protocol, subject pool, outlier handling, and the target perceptual task. Together, these design choices determine what a MOS, DMOS, or preference score can legitimately support. Table~\ref{tab:datasets} compares 3DGS-QA benchmarks by the quality construct and protocol they instantiate.

As shown in Table~\ref{tab:datasets}, GSC-QA \cite{yang2024gscqa}, 3DGS-IEval-15K \cite{xing2025ieval}, 3DGS-VBench \cite{xing2025vbench}, and GScomp-QA \cite{martin2026gscompqa} all conduct quality assessment for compression distortions, but they differ in stimulus design, using single images, fixed video trajectories, model sets, or references to uncompressed Gaussian-splatting content. 3DGS-QA~\cite{wan2026qa} isolates controlled reconstruction conditions and synthetic distortion factors, whereas MUGSQA~\cite{chen2025mugsqa} emphasizes input uncertainty and multi-distance viewing. Aes3D~\cite{xu2026aes3d} further broadens the scope from impairment-focused evaluation to aesthetic assessment, but it should not be treated as a substitute for perceptual quality assessment. Taken together, these benchmarks are complementary, and no single experimental setting can comprehensively characterize 3DGS quality.

In terms of rating protocols, ACR-HR asks for absolute quality with hidden references, DSIS emphasizes impairment relative to a reference, DSCQS collects differential quality under paired presentation, and pairwise comparison estimates preference rather than an absolute MOS. These protocols are all defensible, but they do not answer the same question. For 3DGS, the discrepancy is amplified because attention can shift from texture fidelity in still images to geometry stability during motion and to exploration comfort during free-view navigation.

Stimulus design is equally consequential. Video-based studies approximate continuous viewing and can reveal flicker, blur, and path-dependent artifacts, but a fixed camera trajectory samples only a thin curve through the view space. Image-based studies scale more easily and support fine-grained analysis of training and test views, but they cannot directly measure temporal consistency. Multi-distance protocols better match object inspection because observers naturally zoom in and out, yet they remain pre-rendered rather than interactive. Consequently, the experimental design determines not only measurement noise but also which perceptual dimension is being probed. A concrete example is 3DGS-IEval-15K~\cite{xing2025ieval}. Its image-level design shows that MOS peaks at about 5--6 for training views but only 4--5 for test views, indicating that reference-view selection can systematically overestimate perceived quality.

Despite the diversity in existing 3DGS quality-assessment benchmarks, several central coverage gaps remain. First, free-view interaction is largely absent. Fixed images and videos are repeatable, but they prevent observers from navigating to search for view-dependent failures. Second, temporal and cross-view consistency is only sparsely sampled. A fixed trajectory may reveal flicker or popping, yet it cannot characterize the full quality field around a scene. Third, scene diversity and reconstruction conditions remain uneven, with dynamic content and large-scale environments still underrepresented.

Statistical reporting also remains uneven. Many datasets report the number of observers, the rating protocol, and outlier-handling rules, yet established measures of inter-rater reliability, confidence intervals, and significance testing are not always highlighted. This matters because small differences in correlations with objective quality metrics can be overinterpreted when human judgments are noisy or when artifacts draw attention unevenly across observers. Future datasets should therefore state the intended quality construct explicitly, such as appearance fidelity, method preference, compression impairment, geometry stability, temporal smoothness, or interactive quality of experience.

\begin{table*}[!t]
\caption{Representative subjective 3DGS-QA datasets. Entries summarize the primary quality construct measured by each benchmark.}
\label{tab:datasets}
\centering
\scriptsize
\setlength{\tabcolsep}{1.6pt}
\renewcommand{\arraystretch}{0.98}
\begin{tabularx}{\textwidth}{P{0.105\textwidth} P{0.058\textwidth} P{0.14\textwidth} P{0.175\textwidth} P{0.145\textwidth} P{0.165\textwidth} Y}
\toprule
Dataset & Year & Scene/source & Distortion or focus & Stimulus / protocol & Scale & Main gap \\
\midrule
GSC-QA~\cite{yang2024gscqa} & 2024 & Real, synthetic, dynamic & GGSC compression & Video / DSIS & 120 samples & Single anchor codec; limited distortion coverage. \\
GS-QA~\cite{martin2025gsqa} & 2025 & 8 real scenes & Static GS method comparison & Video / DSCQS & 22 subjects; 64 videos & Small scale; no controlled artifact factors. \\
NVS-QA~\cite{zhang2025nvsqa} & 2025 & 13 dynamic scenes & GS/NeRF method comparison & Video + image & 65 videos + 65 images & Limited scale and no Gaussian-native diagnostics. \\
3DGS-QA~\cite{wan2026qa} & 2026 & Synthetic objects & Controlled reconstruction and synthetic distortions & 6-s video / ACR-HR & 33 subjects; 225 degraded reconstructions & Synthetic-only scenes; controlled but narrow. \\
MUGSQA~\cite{chen2025mugsqa} & 2025 & Synthetic objects & View number, resolution, distance, point-cloud quality & Multi-distance video / crowdsourced & 2,452 subjects; 226,800 raw / 101,555 valid scores & Crowdsourcing and synthetic-source constraints. \\
3DGS-IEval-15K~\cite{xing2025ieval} & 2025 & 10 real scenes & Compression and view-dependent quality & Image / DSIS & 60 subjects; 15,200 images & No temporal or free-view browsing. \\
3DGS-VBench~\cite{xing2025vbench} & 2025 & 11 real scenes & Compression by 6 algorithms & 20-s video / 11-level scale & 50 subjects; 660 models & Fixed paths; no interaction. \\
GScomp-QA~\cite{martin2026gscompqa} & 2026 & 13 real scenes & Compression by 9 GS solutions & Video / reference to uncompressed GS & 20 subjects; 331 video stimuli & Compression-specific; still fixed rendered videos. \\
Aes3D~\cite{xu2026aes3d} & 2026 & 278 reconstructed scenes & Aesthetic quality beyond fidelity & Scene-level aesthetic proxy labels / primitive model & 92,649 views; 8 aesthetic dimensions & Aesthetic rather than impairment QA. \\
\bottomrule
\end{tabularx}
\end{table*}

\section{Objective Metrics and Failure Modes}

To clarify what objective quality assessment can and cannot capture for 3DGS, we first organize existing metrics by the information they observe. Full-reference metrics compare rendered views against ground truth (e.g., PSNR, SSIM, LPIPS, DISTS). No-reference models predict quality directly from rendered images or videos. Conventional 3D quality assessment methods target point clouds or meshes and only partly fit Gaussian primitives. Native 3DGS metrics leverage Gaussian attributes, sometimes combined with rendered evidence.

\begin{figure}[!t]
\centering
\includegraphics[width=0.96\columnwidth]{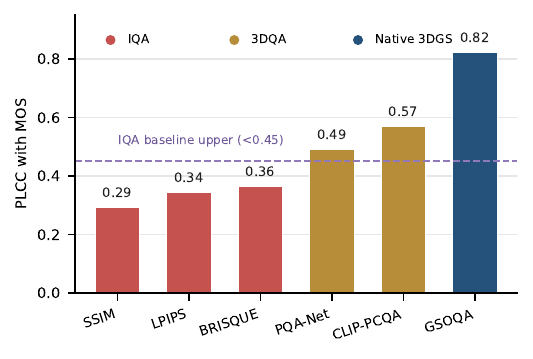}
\caption{Reported PLCC values on the controlled 3DGS-QA benchmark. Inherited IQA and point-cloud-oriented 3DQA baselines remain far below the native GSOQA predictor, indicating that rendered-view or explicit-geometry metrics must be revalidated for primitive-level 3DGS artifacts~\cite{wan2026qa}.}
\label{fig:metrics}
\end{figure}

The failure mode is most evident on the controlled 3DGS-QA benchmark. Conventional rendered-view metrics achieve weak PLCC under 3DGS-specific distortions, whereas a primitive-aware predictor performs substantially better on the same data~\cite{wan2026qa}. Fig.~\ref{fig:metrics} summarizes this PLCC gap between inherited IQA/3DQA baselines and the native GSOQA predictor. This result does not make rendered-image metrics obsolete. It shows that their reliability cannot be assumed based on performance on natural images or average reconstruction benchmarks, and must be validated for the target distortion type and observation protocol.

In practice, metric validity depends on the distortion type and the observation protocol. Under the controlled distortion setting of 3DGS-QA, SSIM, LPIPS, BRISQUE, and point-cloud-oriented baselines show weak correlation with subjective scores~\cite{wan2026qa}. In contrast, on 3DGS-IEval-15K \cite{xing2025ieval}, 3DGS-VBench \cite{xing2025vbench}, and GScomp-QA \cite{martin2026gscompqa}, learned image/video quality models can perform substantially better, because the stimuli and distortions more closely resemble conventional image/video compression under fixed viewing. A metric may therefore align with human ratings under one protocol yet fail under another. The key question is not whether a metric is good in isolation, but which quality dimensions, stimuli, and distortion types justify its use.

Generalizability is also one of the major challenges in objective 3DGS-QA. Learned metrics may overfit to the scene types, distortion patterns, and camera trajectories seen during training. For example, results on the geometry-focused subset of 3DGS-IEval-15K indicate that otherwise strong IQA models degrade when appearance cues are reduced, suggesting that appearance-trained features may underweight structural damage~\cite{xing2025ieval}. Cross-dataset evaluations in NOVA-3DGS provide an even clearer signal that performance can drop markedly when testing on unseen datasets, highlighting limited Generalizability~\cite{piras2025nova}. While NeRF-NQA \cite{qu2024nerfnqa} incorporates view-wise and point-wise reasoning, and more recent PVS-oriented no-reference metric \cite{zhang2026pvs3dmqa} emphasizes temporal video cues, neither direction fully resolves Gaussian-native evaluation. Overall, these findings suggest that current predictors can capture benchmark-specific regularities rather than a stable model of 3DGS perceptual quality.

\section{Toward Native 3DGS Quality Assessment}

A native 3DGS-QA protocol should be explicitly multi-dimensional, because 3DGS impairments span heterogeneous failure modes that are not reducible to a single scalar score. Appearance fidelity remains necessary because users ultimately consume rendered images. Perceptual quality is equally necessary because human preference can diverge from pixel error. Beyond appearance, geometry must be assessed because structural errors may be masked by plausible color. Cross-view consistency is central because quality varies with camera pose. Temporal stability is required to capture popping, flicker, and dynamic-scene artifacts. Interactive QoE becomes relevant whenever users control navigation.

These dimensions are currently evaluated in a disjoint manner. Artifact-mitigation studies isolate individual failure modes. Compression work prioritizes rate–distortion. Subjective datasets rely on heterogeneous stimulus protocols. Metric papers often validate within a single dataset or distortion regime. Consequently, a single MOS-prediction score can obscure which quality dimensions are being traded off. More informative evaluation should consider a structured quality profile rather than a single aggregate. For instance, one method may preserve texture yet introduce unstable floating streaks. Another may reduce primitive count while maintaining temporal stability, and a third may perform well on canonical views but fail under stress views.

View-space specification is the key methodological choice for such a profile. Fixed test views are reproducible but incomplete. Random views improve coverage but can miss rare, high-impact artifacts. User-controlled trajectories are ecologically valid but difficult to standardize. A practical protocol should therefore adopt a layered design that includes canonical views for comparability, held-out views for reconstruction fidelity, targeted stress views for known failure risks, short fixed trajectories for temporal consistency, and limited interactive studies for ecological validation. Although more complex than image-level testing, this design better matches how 3DGS is consumed.

The same layered design should be reflected in reporting. Papers should specify whether evaluated views are training views, held-out views, stress views, fixed trajectories, or user-selected paths. They should provide camera-path metadata and document the renderer configuration, model version, compression settings, and reconstruction conditions. Where feasible, results should be stratified by failure category, since a lower score due to blur is not equivalent to a lower score due to detached geometry. This additional detail increases reporting burden, but it substantially reduces ambiguity about what an objective score or MOS value represents.

Standardization gives the problem practical urgency. MPEG exploration on Gaussian splatting suggests that 3DGS-like representations are entering discussions of interoperable coding and immersive use cases~\cite{mpegGSC,mpeg153}. Compression toolkits likewise emphasize the need for reproducible measurement of rate, rendering cost, geometry fidelity, and resource usage~\cite{splatwizard2025}. Related pressure may also emerge in mobile delivery and asset-exchange ecosystems, including 3GPP-style communication scenarios and glTF-style interchange workflows, even if these tracks have not yet converged on mature QA benchmarks. Moreover, progressive and streaming systems such as LapisGS and PRoGS show that perceived quality can depend on loading order, first-paint time, and bandwidth adaptation, which are not captured by offline frame averages~\cite{shi2024lapisgs,zoomers2025progs}. The risk is that standardization incentives may privilege only easily repeatable rendered-view tests. The opportunity is to define protocols that remain reproducible while still preserving the properties that distinguish 3DGS from a collection of 2D images.

Accordingly, a realistic target is not a universal metric but a modular 3DGS-QA framework. Rendered-view metrics for appearance, primitive-level diagnostics for structural risks, view-space sampling for consistency, temporal probes for stability, and subjective studies for human relevance. Different applications can weight these components differently. Compression, telepresence, XR, editing, and autonomous reconstruction may require different quality profiles, but all require explicit accounting of what each score measures.

\section{Conclusion}
3DGS is increasingly used in real-time rendering, compression, and immersive delivery, yet its quality assessment still largely relies on rendered-view proxies. This survey reviews recent 3DGS-QA work from four perspectives that include representation-specific distortions, subjective benchmark design, objective metric reliability, and requirements for native evaluation. The reviewed evidence shows that current datasets capture different quality constructs, while objective metrics are reliable only under specific distortion sources, stimulus formats, and view and trajectory sampling policies. We organize 3DGS-specific distortions into a representation-aware taxonomy, compare benchmarks along methodological axes beyond scale, and reinterpret objective scores as conditional evidence rather than universal quality indicators. Based on this synthesis, we argue that evaluation should move toward native, multi-dimensional, and reproducible protocols that complement rendered-view metrics with primitive-level diagnostics, view-consistency and temporal probes, and task-aware subjective validation. Benchmarks should state explicitly what is rendered, how it is sampled, what is measured, and which quality construct the score is intended to capture.

\end{document}